\documentclass[]{aa}

\usepackage{graphicx}
\usepackage{txfonts}
\usepackage[colorlinks=true,allcolors=blue]{hyperref}

\usepackage{natbib}
\usepackage{multirow}
\usepackage{multicol}
\usepackage{textcomp}
\usepackage{xcolor}
\usepackage{verbatim}
\usepackage[labelfont=bf]{caption}
\def\MJ{M_\mathrm{J}}
\def\MP{M_\mathrm{p}}
\def\M_sun{M_\mathrm{\odot}}
\def\RJ{R_\mathrm{J}}
\def\RP{R_\mathrm{p}}
\def\Teff{T_{\text{eff}}}
\def\Teffpt{T_{\text{eff},\,pT}}
\def\Xmol{X_{\text{mol}}}
  
\def\XH2O{X_{\text{H}_2\text{O}}}
\def\XCO{X_{\text{CO}}}
\def\XCH4{X_{\text{CH}_4}}
\def\Pb{P_\mathrm{b}}

\DeclareUnicodeCharacter{2212}{-}
\begin{document}

\title{Molecular mapping of the PDS70 system: \\No molecular absorption signatures from the forming planet PDS70\,b\thanks{Based on observations collected at the Paranal Observatory, ESO (Chile). Program ID:0103.C-0680(A)}}

\author{G. Cugno\inst{\ref{ethz}}
\and P. Patapis\inst{\ref{ethz}}
\and T. Stolker\inst{\ref{leiden}}
\and S.~P. Quanz\inst{\ref{ethz}}
\and A. Boehle\inst{\ref{ethz}}
\and H.~J. Hoeijmakers\inst{\ref{lund}}
\and G.-D.~Marleau\inst{\ref{tueb},\ref{bern},\ref{mpia}}
\and P.~Molli\`{e}re\inst{\ref{mpia}}
\and E.~Nasedkin\inst{\ref{mpia}}
\and I.~A.~G.~Snellen\inst{\ref{leiden}}
}

\institute{ETH Zurich, Institute for Particle Physics and Astrophysics, Wolfgang-Pauli-Strasse 27, CH-8093 Zurich, Switzerland\\
\email{gabriele.cugno@phys.ethz.ch}\label{ethz}
\and Leiden Observatory, Leiden University, Niels Bohrweg 2, NL-2333 CA Leiden, The Netherlands\label{leiden}
\and Lund Observatory, Department of Astronomy and Theoretical Physics, Lunds Universitet, Solvegatan 9, SE-222 24 Lund, Sweden\label{lund}
\and Institut f\"ur Astronomie und Astrophysik, Universit\"at T\"{u}bingen, Auf der Morgenstelle 10, D-72076 T\"{u}bingen, Germany\label{tueb}
\and Physikalisches Institut, Universit\"at Bern, Gesellschaftsstr.~6, CH-3012 Bern, Switzerland\label{bern}
\and Max-Planck-Institut f\"ur Astronomie, K\"{o}nigstuhl 17, D-69117 Heidelberg, Germany \label{mpia}
}

\date{Received 22.02.2021 ; accepted 03.06.2021 }

\abstract{}{}{}{}{}
% 5 {} token are mandatory
\abstract
% context heading (optional), leave it empty if necessary
{Determining the chemical properties of the atmosphere of young forming gas giants might shed light on the location in which their formation occurred and the mechanisms involved.}
{We aim to detect molecules in the atmosphere of the young forming companion PDS70\,b by searching for atmospheric absorption features typical of substellar objects.}
{We obtained medium-resolution (R$\approx$5075) spectra of the PDS70 planetary system with the SINFONI integral field spectrograph at the Very Large Telescope. We applied molecular mapping, based on cross-correlation with synthetic spectra, to identify signatures of molecular species in the atmosphere of the planet. }
{Although the planet emission is clearly detected when resampling the data to lower resolution, no molecular species could be identified with the cross-correlation technique. We estimated upper limits on the abundances of H$_2$O, CO and CH$_4$ ($\log(\Xmol)<-4.0$, $-4.1$ and $-4.9$, respectively) assuming a clear atmosphere, and we explored the impact of clouds, which increase the upper limits by a factor up to 0.7 dex. Assuming that the observations directly probe the planet's atmosphere, we found a lack of molecular species compared to other directly imaged companions or field objects. Under the assumption that the planet atmosphere presents similar characteristics to other directly imaged planets, we conclude that a dusty environment surrounds the planet, effectively obscuring any feature generated in its atmosphere. We quantify the extinction necessary to impede the detection ($A_V\approx16-17$~mag), pointing to the possibility of higher optical thickness than previously estimated from other studies. Finally, the non-detection of molecular species conflicts with atmospheric models previously proposed to describe the forming planet.}
{ To unveil how giant planets form, a comprehensive approach that includes constraints from multiple techniques needs to be undertaken. Molecular mapping emerges as an alternative to more classical techniques like SED fitting. Specifically, tuned atmospheric models are likely required to describe faithfully the atmospheres of forming protoplanet, and higher spectral resolution data may reveal molecular absorption lines despite the dusty environment enshrouding PDS70\,b.
}

\keywords{Techniques: imaging spectroscopy -- planets and satellites: formation, atmospheres, individual: PDS\,70}

\titlerunning{Molecular mapping of the PDS70 system}
\maketitle
\definecolor{green2}{rgb}{0,0.8,0.2}
%-----------------------------------------------------------------------

\section{Introduction}
\label{Introduction}
%\vspace{1.cm}
In recent years, high-contrast imaging observations detected a handful of young giant planets on wide orbits \citep[$\beta$\,Pic\,b, 51\,Eri\,b, HIP65426\,b, HR8799\,bcde, HD95086\,b,][]{Lagrange2009, Macintosh2015, Chauvin2017, Marois2008, Rameau2013}. Follow-up observations enabled their characterization and continuously provide new insights on their present atmospheric structure and composition \citep[e.g., ][]{DeRosa2016,Samland2017, Cheetham2019, Gravity2019, Stolker2020}.
At the same time, a lot of effort was invested in the search for young forming planets still embedded in their protoplanetary disks, with several proposed candidates \citep[e.g., LkCa15\,bcd, HD169142\,b, HD100546\,bc, MWC758\,b,][]{KrausIreland2012, Quanz2013, Reggiani2014, Biller2014, Brittain2014, Quanz2015, Sallum2015, Currie2017,Reggiani2018}. However, most recent follow-up studies questioned their existence and were not able to confirm the detections \citep[e.g.,][]{Follette2017, Rameau2017, Ligi2018, Cugno2019a}. 

With two confirmed companions, the PDS70 system \citep{Keppler2018, Muller2018, Haffert2019} offers a unique opportunity to study and characterize protoplanets interacting with their natal environment, sculpting disk gaps \citep{Keppler2019} and accreting material from their surrounding \citep{Haffert2019, Wagner2018, Zhou2021}. 
Follow-up observations resulted in the first attempts to constrain atmospheric parameters and to describe the planets' surrounding. \cite{Wang2020} fitted atmospheric models to determine radius and mass of the companion PDS70\,b ($\RP\sim2-3\RJ$, $\MP\sim2-4\MJ$). \cite{Stolker2020b} obtained comparable results fitting a blackbody to the protoplanet SED, indicating that we are actually seeing radiation reprocessed by the planet environment, which is obscuring our view of strong molecular features. Most recently, \cite{Wang2021} detected molecular absorption features at the blue and red ends of a $K$-band spectrum obtained with VLTI/GRAVITY and found that atmospheric models better fit the planet SED. These latest findings support the tentative detection of H$_2$O feature at 1.4 $\mu$m seen in SPHERE/IFS data \citep{Muller2018}. Moreover, they constrained the mass of PDS70\,b to be $<10\MJ$ in order for the system to be dynamically stable.

\cite{Facchini2021} provided the chemical inventory of the PDS70 disk, revealing molecular emission from 12 molecular species with radial variation indicating dynamic chemistry. The comparison of their data with thermo-chemical models of protoplanetary disks strongly suggests a disk-average C/O ratio > 1. Also, they detected emission from CO in the gap hosting the two planets, indicating that CO is likely currently accreting onto the protoplanets' atmospheres. 

It has been suggested that determining the chemical composition of a planet's atmosphere might play a key role in understanding where, with respect to snowlines, the planet formed \citep{Oberg2011} and to reconstruct the migration history \citep{Madhusudhan2014,Mordasini2016}. Indeed, planet formation history strongly impacts the chemical enrichment of gas giant atmosphere, which depends on the chemical properties of the disk gas from which the planet envelope is accreted.
This, in turn, is affected by disk evolution processes like grain growth, radial drift and ionization \citep{Helling2014, Oberg2016, Eistrup2018,Cridland2019}. Hence, deriving elemental abundance ratios of PDS70\,b, still embedded in its natal environment, offers a unique possibility to verify current chemical modelling of atmosphere enrichment at very early stages. \\

Spectral cross-correlation \citep[e.g.,][]{Sparks&Ford2002} is a widely used technique in exoplanetary science. It was used by \cite{Snellen2014} to measure the rotational velocity of $\beta$\,Pic\,b with the high-resolution spectrograph VLT/CRIRES, and by \cite{Ruffio2019} to estimate the orbital radial velocity of the HR8799\,bc planets. Also, it is regularly used in the analysis of transit spectroscopy datasets \citep[e.g., ][]{Hoeijmakers2015, Giacobbe2021}. 
A similar approach has been used on directly imaged exoplanets to infer the presence of molecules in the atmosphere. Using Keck/OSIRIS data, \cite{Konopacky2013} and \cite{Barman2015} identified spectral features from CO, CH$_4$ and H$_2$O in the spectrum of HR8799\,b and c and used cross-correlation of the planet spectra with molecular absorption templates to successfully confirm the presence of H$_2$O and CO in the planets atmospheres.

Since the spectral signal emitted by the planet is fundamentally different than that of the star, mainly because of the presence of molecules in the atmosphere, medium resolution spectroscopic data can be used to disentangle the planet spectrum from the stellar speckles in the image. \cite{Hoeijmakers2018} adapted the cross-correlation analysis technique to be used not only on measured planet spectra, but also to recognize weak molecular signatures in residuals of medium-resolution spectroscopy data, obtained after removal of the stellar spectrum, instrument systematics and planet pseudo-continuum. They successfully applied molecular mapping to $K$-band $\beta$\,Pic data from VLT/SINFONI, detecting H$_2$O and CO at the planet location with high confidence \citep{Hoeijmakers2018}, demonstrating the efficacy of this approach. Later, \cite{petitditdelaRoche2018} applied molecular mapping on the HR8799\,b planet and \cite{Petrus2020} detected H$_2$O and CO in the atmosphere of HIP65426\,b. \cite{Petrus2020} also compared the atmospheric parameters obtained with the molecular mapping approach with those obtained via matching of synthetic spectra to the data using Bayesian inference. They found that most of the parameter estimates from the two methods agree, confirming that cross-correlation is a powerful tool to investigate giant planet atmospheres. 

In this paper, we analyze for the first time medium resolution spectroscopy data of the PDS70 system, with the goal of searching for signatures of molecular absorption, estimating molecular abundances and elemental abundance ratios to better describe the formation history of the planetary-mass object PDS70\,b. In Sect.~\ref{sec:observations} we describe the data we use in this work. In Sect.~\ref{sec:data_reduction} we detail our data reduction and stellar PSF subtraction. In Sect.~\ref{sec:analysis} we look for molecular signals and report detection limits that we discuss in Sect.~\ref{sec:discussion}. Finally, we summarize our conclusions in Sect.~\ref{sec:conclusions}.

\begin{figure}
    \includegraphics[width=\hsize]{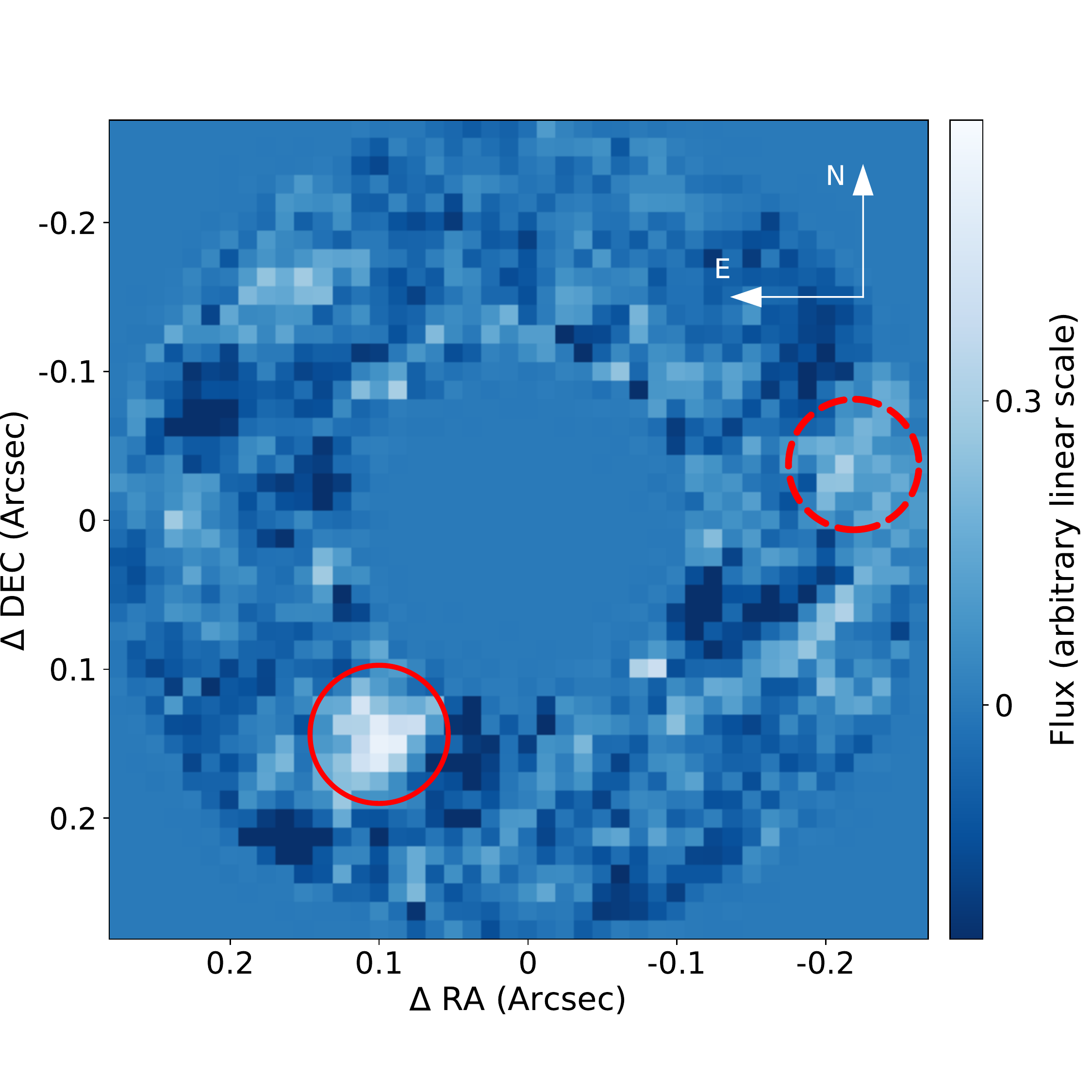}
    \caption{\textit{Left:} PDS70 system with the b planet shown in the red solid circle. Six principal components were removed from the first dataset to detect the planet with S/N~$\sim12$. The dashed circle shows the expected position of PDS70\,c \citep{Mesa2019}.}
    \label{fig:PCA-ADI_white_light}
\end{figure}

%-----------------------------------------------------------------------
\section{Observations}
\label{sec:observations}

We observed the PDS70 system with the SINFONI integral field spectrograph \citep[IFS, ][]{Eisenhauer2003, Bonnet2004} mounted at the European Southern Observatory's Very Large Telescope (VLT) facility in Chile. The observations were carried out in two separate blocks on 2019/05/24 and 2019/06/03. Each observing block consisted of 32 exposures, each with DIT[s]$\times$NDIT=20$\times$8, where DIT is the detector integration time and NDIT the number of DIT repetitions per exposure. Thus, the total time on target each night was $\sim85$ min. In addition to the standard calibration, we took sky frames every two science exposures to sample the background. The data were taken in the $K$-band in the highest spatial resolution mode with a plate scale of $12.5\times25$~mas and a field of view of $0\farcs8\times0\farcs8$. The spectral resolution was $R\approx5075$, measured from the width of OH line spread function.
The images were taken in pupil stabilized mode to enable angular differential imaging \citep[ADI,][]{Marois2008}. The achieved field rotation was 62.4$^\circ$ and 78.0$^\circ$ for the first and the second night, respectively. We dithered the position of the star across the detector to reduce the impact of bad pixels during post-processing. For both nights, the standard star HIP\,81214 was observed for flux and telluric calibration. 
The median DIMM  (Differential Image Motion Monitor) seeing was similar during the two nights, $0\farcs65\pm0\farcs07$ and $0\farcs70\pm0\farcs10$ respectively, while the estimated coherence time $\tau_0$ indicates that the second night had on average less turbulent conditions, with $\tau_0=3.1\pm0.5$ ms and $\tau_0=6.5\pm1.1$ ms respectively. Despite the mean value, we note that both seeing and coherence time indicate that the second night suffered from more variable conditions. Finally, the second observing night had $\sim5$ times lower relative humidity.

\section{Data reduction and quality assessment}
\label{sec:data_reduction}

\subsection{Basic reduction}
\label{sec:basic_reduction}

Both datasets were calibrated using the SINFONI EsoReflex pipeline \citep[version 3.1.1;][]{Abuter2006}, which included dark subtraction, flat fielding, bad pixels removal, wavelength calibration using neon and argon lamps, detector linearity correction and extraction of spectral cubes from the raw frames. The EsoReflex pipeline outputs cubes consistent of 2216 frames, covering a spectral range from 1.929 to 2.472 $\mu$m.

Since images at the edges of the waveband consisted mostly of NaNs and the shorter wavelengths data were dominated by telluric features, we first restricted the spectral range to include frames between 2.088 and 2.440 $\mu$m and removed one row/column at each image side. At this point, only a few NaNs remained in the frames at the location of known bad pixels, which were removed by replacing them with the median of the eight surrounding spatial pixels. 
Across the different wavelengths, the stellar photocenter shifted by up to one pixel within each cube, especially at longer wavelengths. Therefore, a 2D Gaussian function was fitted to each individual frame and to the wavelength-combined (median of the cube along the wavelength axis) frames. Then, images were shifted to a common photocenter given by the wavelength-collapsed images so that the position of the star is the same throughout the whole cube.

\subsection{Stellar PSF subtraction}
\label{sec:PSF-subtraction}

\subsubsection{PCA-ADI approach}
\label{sec:PCA-ADI approach}

After the basic reduction described above, each cube was median-combined along the wavelength axis and we obtained two separate datasets for the observing nights, each of which contained 32 images. The images were cropped to a size of 0\farcs55.

The stellar PSF was removed using {\tt PynPoint}, which uses an algorithm based on Principal Component Analysis (PCA) to reconstruct and subtract the principal modes of the PSF \citep{amaraquanz2012, Stolker2019}. We subtracted 1--15 components and used a central mask of $0\farcs11$ in radius to remove most of the central part of the image without affecting the flux of PDS70\,b. Afterwards, the images were derotated to the same orientation according to the parallactic angle and median-combined. We show the final image obtained for the first dataset in Fig.~\ref{fig:PCA-ADI_white_light}, while planet b is hardly detectable in the second dataset.
We investigated this difference in App.~\ref{App:Data_selection}, concluding that because of more stable weather conditions during the observations the first dataset provides the best chance to study the companion. Thus, in the following we will only use the dataset taken on 2019/05/24. Nevertheless, we verified that the results presented in Sect.~\ref{sec:mol_map_cross_corr} apply to the second dataset and to their combination as well. 
In App.~\ref{sec:broad_band} we treat the wavelength-collapsed cube as an ADI sequence and we investigate the contrast and position of PDS70\,b.
Finally, we note that according to the detection limits it is not possible to detect PDS70\,c in our data. In the images, a point-like emission is somewhat visible at the planet location when selecting the best combination of number of principal components (PCs), image and central mask sizes. However, the signal-to-noise ratio (S/N) remains low ($\lesssim5$) and it is not clear if the emission comes from PDS70\,c or scattered light from the inner disk rim. Therefore, the following analysis focuses on PDS70\,b.

\subsubsection{High-resolution spectral differential imaging (HRSDI)}
\label{sec:mol_map_reduction}

Compared to Sect.~\ref{sec:PCA-ADI approach}, where we looked mainly for the continuum emission, here we want to prepare the data cubes for molecular mapping, meaning that we want to remove the contribution of stellar spectrum from each spatial pixel, while preserving other spectral features like those produced by molecules in a planet atmosphere. Here, the planet continuum is not needed and it can be removed together with the stellar PSF. As in \cite{Hoeijmakers2018}, we used the SkyCalc tool from ESO \citep{Noll2012, Jones2013} to verify if the wavelength solution found by the SINFONI pipeline matches the telluric lines of the synthetic spectrum. We found a mismatch of $\sim60\pm9.8$ km/s ($\sim$ 2 pix) between the wavelength solution delivered by the pipeline and the sky simulations provided by SkyCalc. We corrected this effect by shifting the wavelength of each frame to the rest frame given by the SkyCalc spectrum.

After the basic reduction explained in Sect.~\ref{sec:basic_reduction}, the PSF is removed using a spectral differential technique based on \cite{Hoeijmakers2018} and \cite{Haffert2019}. Each spatial pixel is divided by its total flux after 12$\sigma$ clipping to remove remaining hot pixels, and all the resulting spectra are median-combined to generate a reference spectrum, which does not include any planet signal because of the use of the median. The spectrum of each spatial pixel is then divided by this reference spectrum. The residuals are low-pass filtered using a Savitky--Golay filter \citep{Savitzky1964} of order 3 with a window of 31 channels ($\sim\,7.6$~nm) to remove spectral features as broad or broader than the filter size, leaving only high-frequency features. The filtered spectrum is then multiplied again by the reference spectrum. Modelling $\lambda$-dependencies over the PSF shape, this process delivers a continuum spectrum of the stellar PSF at each spatial pixel that does not include spectral features coming from the planet and the star. The final residuals are given by the subtraction of the initial spectrum from the local reference filtered spectrum. The spatial pixels of each individual cube are then used for building a reference library through PCA to remove correlated noise within the cube itself. This step is performed cube by cube and each time the first 20 components were removed. 

At this point, correlated (instrumental) noise between cubes still dominates the central part of the images and high-frequency residuals that remained in the data could make the detection of molecular signatures from planets still difficult. In particular, (i) telluric residuals could mimic the detection of molecular lines (especially H$_2$O) at any location in the image, and (ii) H$_2$O and CO emission lines from the inner unresolved circumstellar disk could be detected with cross-correlation (see Sect.~\ref{sec:mol_map_cross_corr}). Those features could have counterbalanced absorption features from the planet atmospheres of the same molecules, impeding their identification. We found that applying PCA a second time is the best method to remove these last noise sources. This time, we used the entire dataset (32 cubes) to build the reference library, and we removed two principal components.

After removing the stellar spectrum from each pixel, the star was centered, the cubes were derotated to a common field orientation using their parallactic angles, and the frames corresponding to the same wavelength were median combined, resulting in a residuals cube made of 1479 frames. Finally, a mask of radius $0\farcs11$ was applied at the center of each frame. 

\section{Analysis and Results}
\label{sec:analysis}

\begin{figure*}
    \includegraphics[width=1.\hsize]{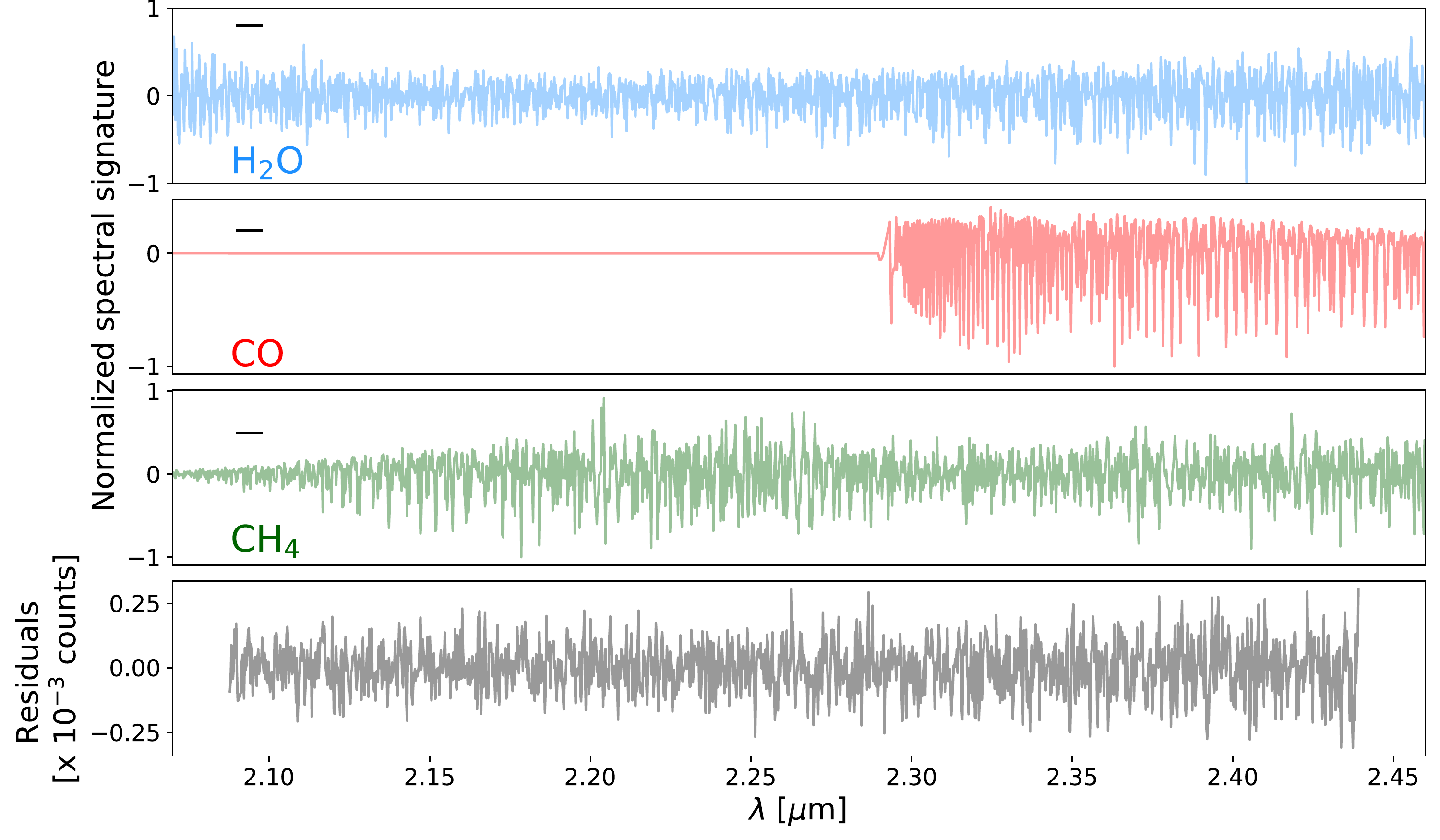}
    \caption{Molecular templates after removal of the continua (H$_2$O in blue, CO in red and CH$_4$ in green) and residuals of the data at the position of the planet (grey). No broad feature is visible and only individual absorption lines at the instrument resolution are cross-correlated. The small black dashes in the top left of the first three panels represent the filter size for continuum removal. Features smaller than this are kept in the residuals, while larger structures are removed as part of the continuum.
    }
    \label{fig:Spectral_templates}
\end{figure*}

In this section, we exploit the $R\approx5075$ resolution of the SINFONI spectrograph in order to detect and, if possible, quantify the presence of molecules in the atmosphere of PDS70\,b. To do that, we cross-correlate the residuals of the data with templates containing characteristic absorption features from specific molecules detected in the atmosphere of low-mass objects.

\begin{figure*}
    \includegraphics[width=1.\hsize]{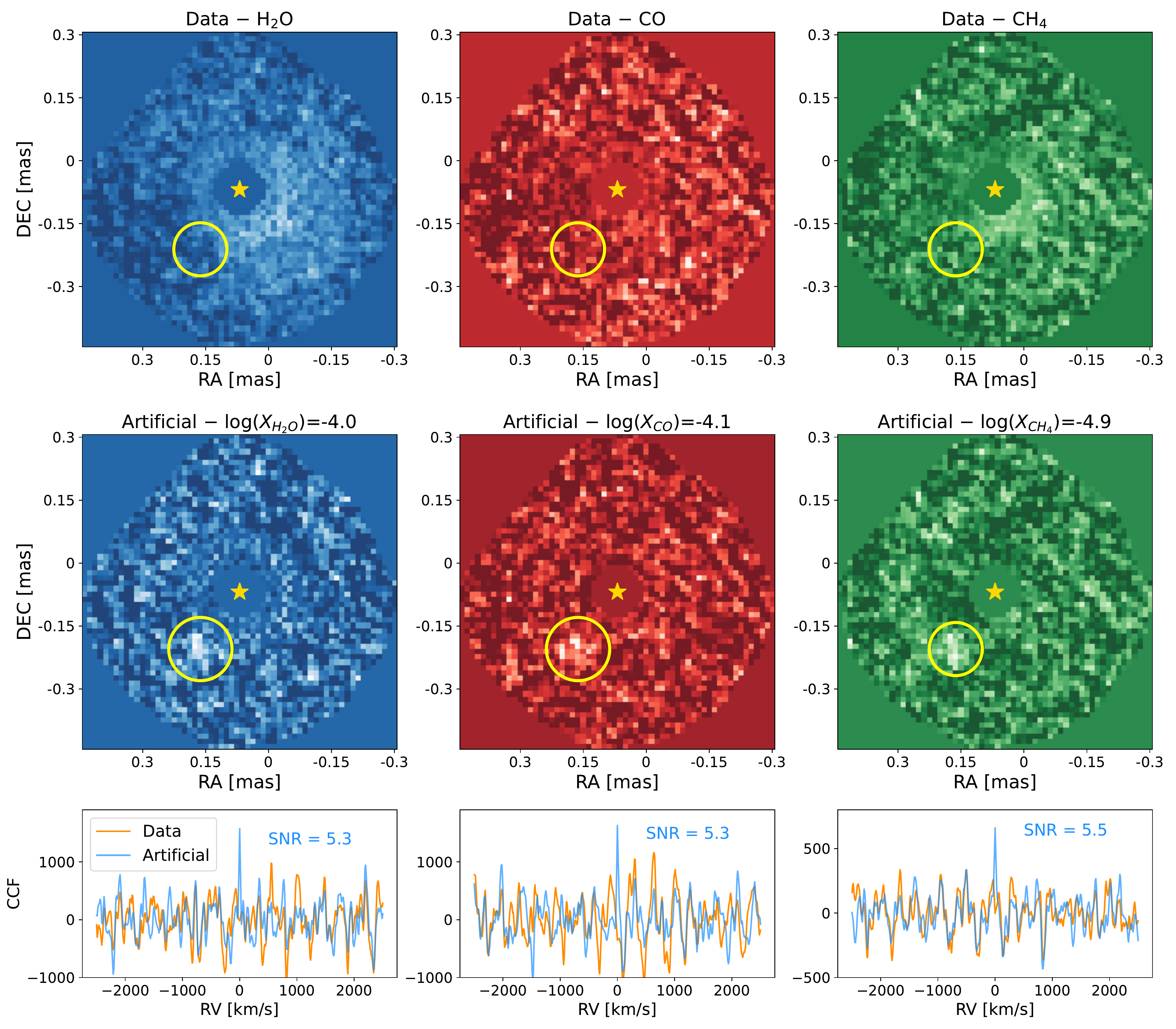}
    \caption{\textit{Top row:} H$_2$O, CO and CH$_4$ molecular maps of the PDS70 system at RV$\sim-5$ km/s. The central star represents the position of PDS70\,A, while the yellow circle indicates where a signal from PDS70\,b should be found. \textit{Middle row:} Same as in the top row, but a planet with the molecular abundances indicated on the top of each panel has been inserted in the data and detected with S/N$\sim5$. \textit{Bottom row:} CCF as a function of the radial velocity at the location of the planet. The orange lines represent for each molecule the CCF of the data, while the blue ones the CCF of the inserted artificial planets.}
    \label{fig:mol_map}
\end{figure*}

\subsection{Spectral templates}
\label{sec:spectral_templates}
In order to use cross-correlation and identify signatures from individual molecules, we generated templates with {\tt petitRADTRANS} \citep{Molliere2019}, a Python package for calculating transmission and emission spectra of exoplanets.
We used the thermal structure grids calculated with {\tt petitCODE} \citep{Molliere2015, Molliere2017} provided in \cite{Samland2017} for objects with an effective temperature of $\approx 1200$ K \citep{Stolker2020b}, 
solar metallicity \citep[as assumed for PDS70A by][]{ Muller2018} and $\log(g) =3.0$. The surface gravity was chosen to be 3.0 as it was the closest available value assuming $M_{pl}\sim1.5\,M_J$ and $R_{pl}=3.0$ from \cite{Stolker2020b}.
Using the vertical thermal profile we created high-resolution emission spectra of planets, with an atmosphere where only one molecular species (H$_2$O, CO or CH$_4$) is present.
The mass fraction of the complex molecule was arbitrarily chosen based on two criteria: it must create well defined molecular absorption lines without saturation, and it must be compatible with mass fractions previously estimated in low mass objects \citep[e.g.][]{Lavie2017}. These were chosen to be $\log(\XH2O)=-2.7$, $\log(\XCO)=-2.5$, and $\log(\XCH4)=-4.0$, where $X$ represents the mass fraction of the molecule. The mass fractions of H$_2$ and He were set to 75\% and 25\% of the remaining atmosphere mass \citep{Line2012}.  We focused on these molecules as chemical equilibrium models predict them to be the most abundant molecules with significant features in the observed wavelength range at this temperature \citep[see Fig.~11 of][]{Stolker2020}. The thermal structure was not adapted to the molecular opacities as detections mostly depend on the position of absorption lines, which, unlike line depths, does not change when altering the $p$--$T$ profile of the atmosphere. Relative abundances were kept constant with altitude \citep[e.g.,][]{Todorov2016, Line2017}. In the calculation, H$_2$--H$_2$ and H$_2$--He continuum opacities were included, while clouds were not considered.  To first order, the presence of clouds would only lower the amplitude of the spectral features, therefore scaling the cross-correlation signal without impacting the S/N of potential detections.

The resulting simulated spectra were resampled to the spectral resolution of the data, and a Savitzky--Golay filter \citep{Savitzky1964} with window of 31 channels ($\Delta\lambda \approx 7.6$~nm) and order 3 was applied to approximate the continuum emission. The latter was subtracted from the original spectrum in order to obtain the spectral signature of the molecules without low-frequencies contributions from the continuum. The molecular templates, together with the residuals that were cross-correlated with the data, are shown in Fig.~\ref{fig:Spectral_templates}.

\subsection{Cross-correlation}
\label{sec:mol_map_cross_corr}
To search for signals due to molecular absorption from H$_2$O, CO and CH$_4$, each spatial pixel in the residual cube is cross-correlated with each of the molecular templates over a range of radial velocities between $-2500$~km/s to 2500~km/s in steps of~10 km/s. For each of the templates, this process resulted in a new cube made of 501 images, where each image represents the cross-correlation function (CCF) at a different radial velocity \citep{Hoeijmakers2018}. The more similar the model template is to the real emission spectrum of the planet, the larger the value of the CCF at the location and radial velocity of the planet will be.

To verify that our methods are effective in detecting molecules, we downloaded archival data of $\beta$\,Pic taken in the same mode as our PDS70 data and reduced them looking for H$_2$O and CO signals as presented in \cite{Hoeijmakers2018}. We successfully identified the presence of both molecules, as shown in App.~\ref{sec:betaPicb}.

Once the pipeline was verified, we searched for peaks in the CCF at the location of PDS70\,b found in App.~\ref{sec:broad_band} at radial velocities of RV$_{\rm pl}$=RV$_{\rm orbit}$+RV$_{\rm bary}=4.3-9.9\approx-5.6$ km/s, where the first term is the radial velocity component of the keplerian velocity with respect to the system velocity \citep{Haffert2019} and the second one represents the barycentric velocity during the observation calculated with the \texttt{PyAstronomy} python package\footnote{\url{https://pyastronomy.readthedocs.io/en/latest}}. Given the instrument resolution, each channel corresponds to a shift in radial velocity of $\sim30$ km/s, indicating that a higher precision in the determination of the planet radial velocity is not necessary. No signal was detected for the three molecules over the full RV range. The panels in the top row of Fig.~\ref{fig:mol_map} show the molecular maps at RV$_{\rm pl}$ for H$_2$O, CO and CH$_4$, respectively, while the CCFs at the planet position are shown in orange in the bottom row of the same figure for the three molecules.

In addition to the pixel-by-pixel approach explained above, we summed the residual flux at each channel in a circular aperture of $0\farcs0275$ in radius placed at the planet location to boost the signal to noise ratio of molecular features. No significant peak in the CCF was detected.

\subsection{Detection Limits}
\label{sec:mol_map_det_lim}
To assess the quality of the data and the scientific relevance of the non-detection of molecules in the photosphere of PDS70\,b, we estimated limits on molecular abundances by inserting artificial companions in the data before subtracting the PSF and looking for the lowest molecular abundance that could be detected with our method. To do that, we fixed the thermal structure of the atmosphere to self-consistently calculated $p$--$T$ profiles as we lack empirical constraints on the protoplanet's thermal structure. Under this assumptions, determining precise abundance limits from emission spectra is a difficult task, as changing abundances of the main absorbers theoretically involves major changes in the $p$--$T$ structure and the emitted SED that could not be taken into account.

Planet spectra were generated with {\tt petitRADTRANS}, with the thermal structure for $\Teff=1200$~K, $\log(g)=3.0$ from \cite{Samland2017}. We varied $\Xmol$, the mass fraction of the molecule we are looking for while we kept the amount of H$_2$ and He constant in the rest of the atmosphere (75\% H$_2$, 25\% He). The planet was placed at a distance of 113.4 pc \citep{Gaia2018} and the planet radius was assumed to be 3.0 $\RJ$ \citep{Wang2020, Stolker2020b}. Once generated, the planet emission spectrum was Doppler-shifted by $-5$ km/s and down-sampled to the spectral resolution of the data. Then, the contrast between the injected signal and the stellar flux (estimated in App.~\ref{sec:flux_calibration}) was calculated.
Finally, in each frame a copy of the stellar PSF rescaled by the contrast was inserted in the pre-PSF-subtracted cubes at the position of PDS70\,b from App.~\ref{sec:broad_band}. The new data with the inserted planets were reduced with the same pipeline detailed in Sect.~\ref{sec:mol_map_reduction}, and the same templates from Sect.\,\ref{sec:mol_map_cross_corr} were cross-correlated with the residuals to find molecular absorption features. 

We estimated the significance of the detection calculating its S/N as the ratio between the peak value of the CCF and the standard deviation of its remaining values at least 20 RV steps (i.e., 200 km/s) apart from the peak. To correct for autocorrelation effects on the noise measurement, particularly strong in the case of CO, we calculated the autocorrelation function of the templates, rescaled it so that its peak is at the same level as the data-template CCF peak and then subtracted it from the original CCF for |RV|\,>\,200 km/s. This way, the autocorrelation signal is not counted in estimating the noise. Finally, we adjusted $X_{\text{mol}}$ to obtain S/N$\approx5$.

Within our assumptions on the atmospheric parameters, the VLT/SINFONI data exclude $\log(\XH2O)>-4.0$, $\log(\XCO)>-4.1$ and $\log(\XCH4)>-4.9$, because molecular mass fractions larger than these would have been detected by our methods with S/N>5. The middle row of Fig.~\ref{fig:mol_map} shows the molecular maps of PDS70 in H$_2$O, CO and CH$_4$ at the RV of the artificial planet for the limit mass fractions provided above. The signal of the three molecules is detected with S/N$\,\gtrsim\,$5 as shown with blue lines in the CCF of the bottom row panels. The non-detections of H$_2$O and the low abundance limits we calculated within the framework of our assumptions appear to be in disagreement with the GRAVITY spectrum \citep{Wang2021}, where the presence of H$_2$O in the atmosphere was inferred thanks to the spectral slope at the red end of the spectrum. We address this possible inconsistency in Sect.~\ref{sec:comparison_previous_studies}.

In order to account for systematic wavelength calibration effects from the EsoReflex pipeline we tested the impact of a range of errors on the cross-correlation signal. Using the known spectral position of telluric lines in the data we estimate the mean error ($\sigma=8.4$~km/s) of the wavelength calibration across the field of view. We proceed to distort the wavelength coordinates of a model spectrum by adding a sine function as a systematic error. The same non-distorted spectrum is used as the template and is cross-correlated with the distorted spectrum. We vary both the frequency and amplitude of the sine function, constrained by the size of the spectrum array (the largest frequency being a single wavelength bin and the smallest the size of the array) and measured wavelength calibration error respectively. The H$_2$O and CO templates were tested individually since wavelength calibration errors might have different impact on each molecule. We found a loss in SNR of maximum 20\% in the extreme case. Thus, the molecular non-detection can not be explained with instrumental calibration effects

\begin{figure*}
    %\centering
    \includegraphics[width=\hsize]{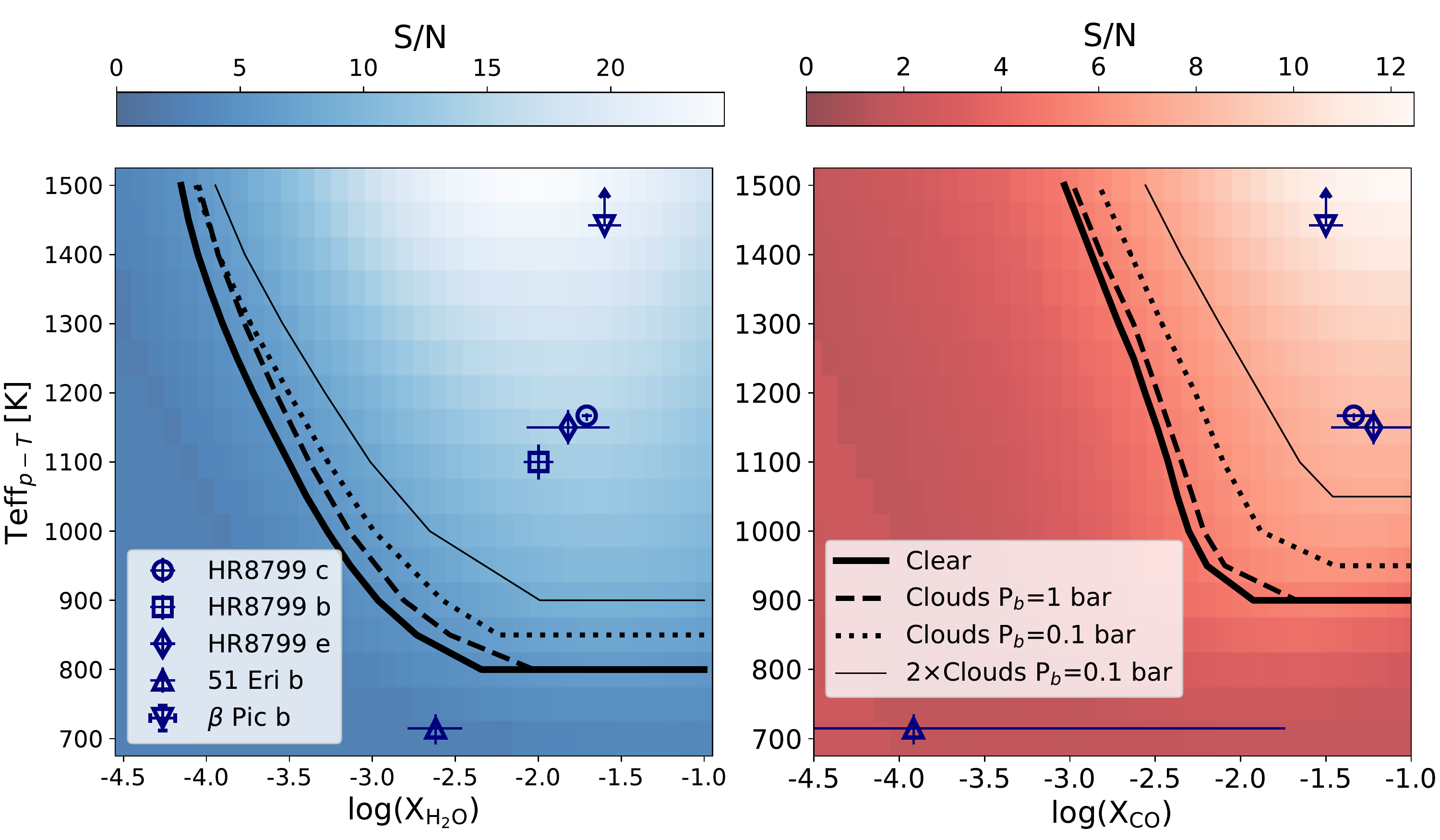}
    \caption{H$_2$O (left) and CO (right) detection maps for clear atmospheres as a function of mass fraction $\Xmol$ and effective temperature $\Teffpt$ of the planet. Markers represent measurements for other directly imaged companions. Thick solid lines separate the region with S/N>5 from regions with S/N<5, i.e., represent the molecular detection limit as a function of planet temperature. Dashed, dotted and thin solid lines represent the molecular detection limits for different cloud configurations based on \cite{Molliere2020} as described in the main text ($\Pb=10^{-3}$~bar is not shown). PDS70\,b's temperature ranges between 1200 and 1400 K \citep{Stolker2020b, Wang2021}.  The effective temperature of $\beta$\,Pic\,b is well constrained as $\Teff=1742\pm16$~K \citep{Gravity2020} and the arrow in its marker indicates that it is higher than the maximum $\Teffpt$ considered here and should not be interpreted as a lower limit.}
    \label{fig:teff_abund}
\end{figure*}

\section{Discussion}
\label{sec:discussion}
As chemical equilibrium models predict very low abundances of methane for objects with $\Teff \approx 1200$ K \citep[e.g.,][]{Zahnle2014}, we focus the upcoming discussion on H$_2$O and CO only, as those are the more common molecules to be expected in self-luminous objects according to chemical models \citep{Zahnle2014} and existing observations \citep{Hoeijmakers2018, petitditdelaRoche2018, Petrus2020, Miles2020}. 

\subsection{Temperature dependence}
\label{sec:temp_dependence}
The uncertain physics of PDS70\,b prevent a clear determination of the thermal structure of the planet. As atmospheric chemistry is highly dependent on the temperature \citep[e.g.,][]{Zahnle2014}, we evaluate the impact of varying the planet's assumed temperature on the upper limits of the molecular abundances.
We estimated the S/N of molecular detections for effective temperatures in the range $\Teffpt=700$--1500~K (in steps of 50~K), with $\log(\Xmol)$ between $-4.5$ and $-1.0$~dex (in steps of 0.1~dex). Here we denote by $\Teffpt$ the effective temperature used to calculate a given $p$--$T$ profile (keeping $\log(g)=3.0$ fixed). The generated spectra have a different $\Teff$ because they are calculated with different abundances. However, for the $\Teffpt=1200$~K case, the $K$-band spectrum of the inserted planet agrees with the flux measured by SPHERE/IRDIS for PDS70\,b \citep{Muller2018}, while $\Teffpt=1500$~K implies a $\sim50\%$ brighter planet. An overall brighter planet would induce a stronger signal in the CCF and a given spectral signature would be easier to detect. Hence, we conclude that the flux inconsistencies due to the choice of the thermal structure may lead to minor underestimate of the molecular abundance limits of the order of 0.3~dex at most, decreasing when nearing $\Teffpt=1200$~K. This value has been calculated by varying the planet radius in order to match the SPHERE/IRDIS measured $K$-band flux.

Furthermore, realistic emission spectra are likely the result of absorption features produced by multiple molecular species at different altitudes, thus polluting the signal of each individual molecule. Therefore, when generating the planet spectrum with varying $\Xmol$, we include also the amount of H$_2$O/CO found by \cite{Wang2020_HR8799} in HR8799\,c ($\Teff\approx1200$ K, similar to PDS70\,b). 

In the same way described in Sect.~\ref{sec:mol_map_det_lim}, artificial planets were inserted in the data and we searched for molecular signatures calculating the S/N. The resulting charts of S/N as a function of $\Xmol$ are shown in Fig.~\ref{fig:teff_abund} for H$_2$O and CO. The solid black lines show where S/N = 5.
Our data are more sensitive for hotter planets, where in case of a clear atmosphere we expect to be able to detect mass fractions down to $\sim10^{-3.6}$. The difference between this value and the one presented in Sect.~\ref{sec:mol_map_det_lim} and Fig.~\ref{fig:mol_map} is due to the polluting effect of the other molecule in the emission spectrum. 
At lower temperatures, molecular mapping is less effective for two main reasons: (i) cooler planets are fainter and produce weaker spectral signals, and (ii) absorption lines are less prominent, as the difference in temperature between atmospheric layers is smaller and the thermal structure tends to be more isothermal, especially at low $\Teff$. Below 850 K, we are not sensitive to these molecules. This is not necessarily the case for fully self-consistent temperature structures, as changing $\XCO$ and $\XH2O$ result in the $p$--$T$ profile to be adapted.

For comparison, Fig.~\ref{fig:teff_abund} also reports the estimated abundances of the directly imaged companions with available measurements in the evaluated temperature range. Abundances for HR8799\,b, HR8799\,e and $\beta$\,Pic\,b planets are taken from \cite{Lavie2017}, \cite{Molliere2020} and \cite{Gravity2020}. For HR8799\,c we used \cite{Wang2020_HR8799}, who used atmospheric retrievals on Subaru/CHARIS, Gemini/GPI and Keck/OSIRIS data to constrain atmospheric properties of the planet. 
For 51\,Eri\,b, we used the retrieval analysis performed by \citet[submitted]{Whiteford2021}. We note that our method would not have been able to find molecules in the atmosphere of 51 Eri b, the main reason being the much lower effective temperature \citep[][submitted]{Whiteford2021}. 

We note that Fig.~\ref{fig:teff_abund} strongly depends on the temperature profile of the planetary atmosphere. Indeed, a profile with a smaller temperature gradient between inner and outer layers reduces the molecular absorption signature and therefore our ability to detect such molecules. Pushing this consideration to its limit, an isothermal profile causes the atmosphere to be featureless, resulting in a black body emission. Under the assumption of a clear atmosphere, Fig.~\ref{fig:teff_abund} suggests that the chemistry of PDS70\,b is different than the one of other directly imaged companions with similar effective temperature. Furthermore, the non-detection of CO in the planet atmosphere conflicts with the detection of CO gas in the disk gap harbouring the protoplanet, as the latter is likely directly accreted onto the young planet \citep{Facchini2021}. This inconsistency brings us to consider other sources of higher opacity.

\subsection{Clouds}
\label{sec:clouds}
Clouds have been found to be key elements to describe the atmosphere of many gas giant planets \citep[e.g.,][]{Madhusudhan2014, Molliere2020, Gao2021}. Their presence causes the molecular features to be damped, attenuating their signature in the planet spectra and shifting the SED to the red. Therefore, the abundance limits calculated in Sect.\,\ref{sec:mol_map_det_lim} and Sect.~\ref{sec:temp_dependence} are not valid when clouds are included. Although their relevance in describing gas giants atmosphere is undisputed, the details of cloud physics remain difficult to describe and parametrize, and uncertainties and assumptions still dominate their treatment. Hence, estimating upper limits on the abundances in the presence of clouds would cause the parameter space to be too large to be explored. Instead, we undertake a qualitative approach, in which we present the impact that four concrete cloud configurations have on the molecular mass fraction upper limits. The cloud parametrization is based on the findings of \cite{Molliere2020} for the planet HR8799\,e, which has $T_{\text{eff}}=1154\pm50$ K, a value similar to that of PDS70\,b \citep{Stolker2020}. Briefly, in \cite{Molliere2020} five cloud parameters were free to explore the parameter space: mass fraction of Fe and MgSiO$_3$ at the cloud base compared to their equilibrium chemistry value, settling parameter $f_{\rm sed}$, eddy diffusion coefficient $K_{zz}$ and width of the log-normal size distribution of the particles $\sigma_g$. For each of those parameters, we used the value reported in Fig.~4 of \cite{Molliere2020}, and we used Eq. 3 from the same paper to describe cloud mass fractions as a function of the altitude. The cloud base pressure $\Pb$ was manually set to 1, 0.1 or $10^{-3}$~bar for both cloud species to obtain three of the four investigated scenarios. The altitude of the clouds in the planet atmosphere strongly depends on the temperature structure. Exploring the parameter space for $\Pb$ therefore allows us to consider the impact of different clouds independent from the assumed thermal structure. Finally, the fourth cloud scenario is obtained multiplying the Fe and MgSiO$_3$ mass fractions at $\Pb=0.1$~bar with a factor of two, i.e. doubling the cloud's optical thickness.

We repeated the calculation from Sect.~\ref{sec:temp_dependence} and we report the lines at S/N=5 in Fig.~\ref{fig:teff_abund}. The case with $\Pb=10^{-3}$~bar is not shown, as clouds at high altitudes present low particle densities, unable to affect the molecular spectral features. Lines for this scenario would overlap the solid lines for clear atmospheres. As expected, as a consequence of clouds, the upper limits we derived are higher, meaning that only larger molecular abundances could have been detected. These values are only valid for this specific realization and they would be different for different parametrizations. Nevertheless, they give a sense of the impact that clouds might have on the detectability of molecules in planet atmospheres. Again, molecular mapping is very sensitive to the presence of H$_2$O and CO for hot objects. Indeed, under these assumptions, our data strongly suggests that much higher optical depths than what were used above are necessary to hide the spectral features produced by molecules, and even clouds may not be sufficient to explain the non detections in the SINFONI data, assuming PDS70\,b's chemistry to be similar to the one of other directly imaged objects.

\begin{figure}
    \centering
    \includegraphics[width=0.5\textwidth]{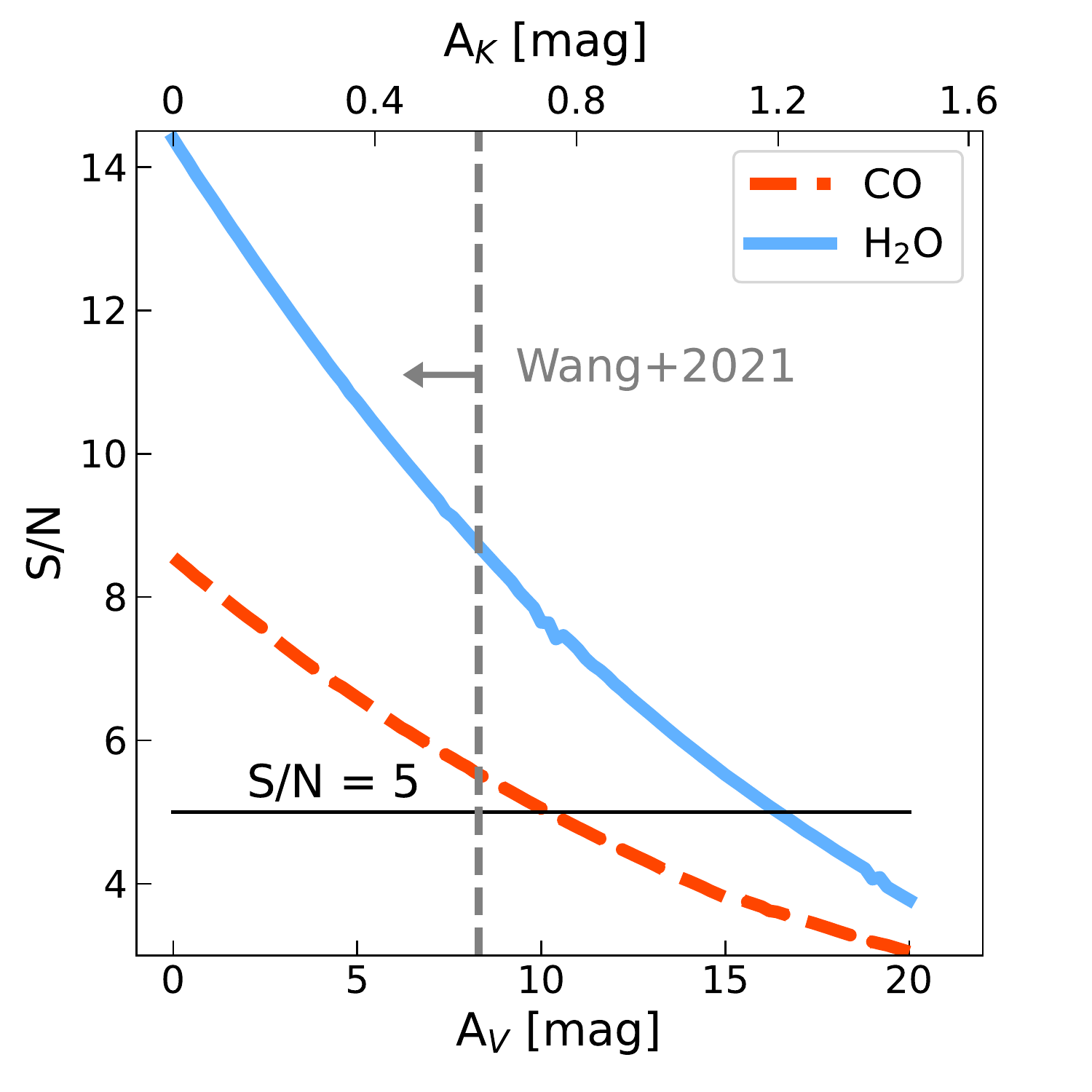}
    \caption{Signal to noise ratio of the detection of H$_2$O and CO as a function of extinction affecting the emission spectrum of the planet. Assuming PDS70\,b has the same chemical composition as HR8799\,c, an additional extinction with $A_V\approx16-17$ mag ($A_K\approx1.2$ mag, using the extinction law from \citealp{WangChen2019}) is required in order not to detect molecules with S/N>5. The vertical dashed line shows the maximum value obtained by SED fitting from \cite{Wang2021}. Such an extinction would have revealed molecules in the planet atmosphere with S/N$\sim$7-9.}
    \label{fig:extinction}
\end{figure}

\subsection{Extinction}
\label{sec:extinction}
Assuming a composition for the planet similar to what has been measured for other gas giants, we can estimate the minimum extinction necessary to impede the detection of the absorption spectrum with molecular mapping. Following \cite{Wang2021}, we used the near-infrared ($1.0-3.3~\mu$m) extinction law derived by \cite{WangChen2019} to attenuate the emitted spectrum, with $A_\lambda\propto\lambda^{-2.07}$. 
We applied an extinction factor $10^{-A_\lambda/2.5}$ to the generated planet spectrum, where $A_\lambda$ is the extinction to the radiation emitted by a planetary atmosphere at wavelength $\lambda$. Applying such an absorbing element causes the line features to be less prominent in the residuals (although the line contrast relative to the planet continuum remains the same) and therefore the molecules harder to be detected. We benchmarked again our data against the molecular abundances from \cite{Wang2020_HR8799} for HR8799\,c. We applied different levels of visual extinction $A_V$ to its emission spectrum, from $0$~mag to 20~mag in steps of 0.2~mag, using \cite{WangChen2019} to calculate the extinction in the $K$-band. For each of the assumed values we estimated the S/N of the detection of H$_2$O and CO. These are shown in Fig.~\ref{fig:extinction}. Assuming the HR8799\,c chemical composition, our data indicates that an extinction of $A_V\approx16-17$ mag ($A_K\approx1.2$ mag) is required to attenuate molecular signatures of H$_2$O in the object's emission spectrum, while a lower extinction ($A_V\sim10$~mag) is necessary to explain the non detection of CO. These values, particularly the one for H$_2$O, are much larger than the largest value from \cite{Wang2021}, which is $A_V=8.3$ mag when coupled with Exo-REM atmospheric models \citep{Charnay2018}.
Assuming only one component for the flux, our analysis suggests that the intrinsic planet emission would need to be a factor $\sim3$ larger than what is measured in the $K$-band in Sect. \ref{sec:PCA-ADI approach} and in previous observations \citep{Muller2018, Christiaens2019_CPD, Wang2021}. A larger radius could be invoked to explain the larger intrinsic flux, but this would in turn increase again the amplitude of molecular features in the spectrum, which is inconsistent with non-detection via molecular mapping. To attenuate the absorption signature while maintaining the flux at the observed level, additional radiation reprocessed by the planet dusty environment, likely coming from both the planet interior and the accretion shock, needs to be invoked \citep{Stolker2020b, Wang2021}. Following \cite{Stolker2020b}, we can use $\dot{M}=5\times10^{-7}\MJ/\mathrm{yr}$, $M_p=1\MJ$ and $R_p=2.0\,\RJ$ to estimate the accretion luminosity $L_{\mathrm{acc}}=G\,M_p\,\dot{M}/R_p \simeq 7.0\times10^{-5}\,L_\odot$. If we consider the accretion process to be fully thermalized, we expect its reprocessed emission to appear in form of a black-body. To first order, we can use the photometric radius $R_\mathrm{phot}=3.0 \,\RJ$ from \cite{Stolker2020b} to constrain its effective temperature ($\sim950$ K) and its contribution to the $K$-band flux ($\sim30\%$ of the flux measured by GRAVITY, \citealt{Wang2021}). We note that \cite{Stolker2020b} used the lower limit of the mass accretion rate measured by \cite{Hashimoto2020}, and that using larger mass accretion rates increases $L_\mathrm{acc}$ and in turn the $K$-band contribution. A more sophisticated radiative transfer modeling will be crucial to investigate this scenario.

At this point, it is not possible to conclude where the dust responsible for such a high extinction is located, meaning that it could be in form of (i) very thick atmospheric clouds as we addressed in Sect.~\ref{sec:clouds}, (ii) external to the planet in form of a dusty shell enshrouding it, or (iii) due to circumstellar disk material. Of course, a combination of these elements is possible.
The presence of circumstellar disk material depends on the mass of PDS70\,b. Indeed, more massive objects are expected to carve larger gaps and clear them from circumstellar material, therefore lowering extinction effects due to the protoplanetary disk \citep[see, e.g.,][]{Sanchis2020, Szulagyi2020}. A mass in the $1-4\,\MJ$ range was inferred for PDS70\,b \citep{Wang2020, Stolker2020b, Wang2021}, which might not be high enough to completely clear the protoplanetary disk gap. However, the presence of a second companion in the same cavity and at larger separations, together with the dust being trapped in the outer ring \citep{Keppler2019}, could prevent replenishment of new material and maintain the gap clear.

Conversely, when adopting the estimated $A_V$ from \cite{Wang2021} ($A_V\sim5-8$ mag), we conclude that the molecular abundances in the atmosphere of PDS70\,b are lower than what we observed in more mature gas giant planets of similar temperature. Precise measurements on a sufficiently large number of planetary mass objects is required in order to further investigate this possibility.

The absence of absorption features together with the possibility of a very high extinction as indicated by our analysis could suggest a scenario similar to Fomalhaut\,b \citep{Kalas2008}, where scattered light is detected from a point-like source located at $\sim$18 AU. As Fomalhaut\,b emission was only detected at optical wavelengths, and not in thermal emission, several ideas were proposed to explain the observations, some involving a planetary object. Examples are a low-mass planet surrounded by a ring system \citep{Kalas2008} or by a swarm of collisional satellites \citep{Kennedy&Wyatt2011}, but also a transient dispersing dust cloud \citep[e.g.,][]{Janson2012, Gaspar&Rieke2020} or a background object \citep{Neuhauser2015}. However, despite the similarities, several empirical observations of PDS70\,b seem to discard this hypothesis. First, a scattered light scenario can not explain the dip at the end of the GRAVITY spectrum \citep{Wang2021}. Second, the H$\alpha$ line emission measured in \cite{Haffert2019} appears to be shifted with respect to the expected radial velocity if it were scattered light from material moving with keplerian motion, and its width is narrower than the one from PDS70A. Third, it is possible that PDS70\,b is located in the shadow of the inner disk resolved by \cite{Wang2021}, thus it might be only weakly irradiated from the star. Finally, in reflected light the protoplanet's SED is expected to follow the stellar SED. However, the measured flux at short wavelengths decreases rapidly below the detection limit in the Y band \citep{Muller2018}.  In summary, although scattered light could contribute to its emission spectrum, it seems unlikely that this is the dominant source of radiation for PDS70\,b.

\begin{figure}
    \includegraphics[width=1.\hsize]{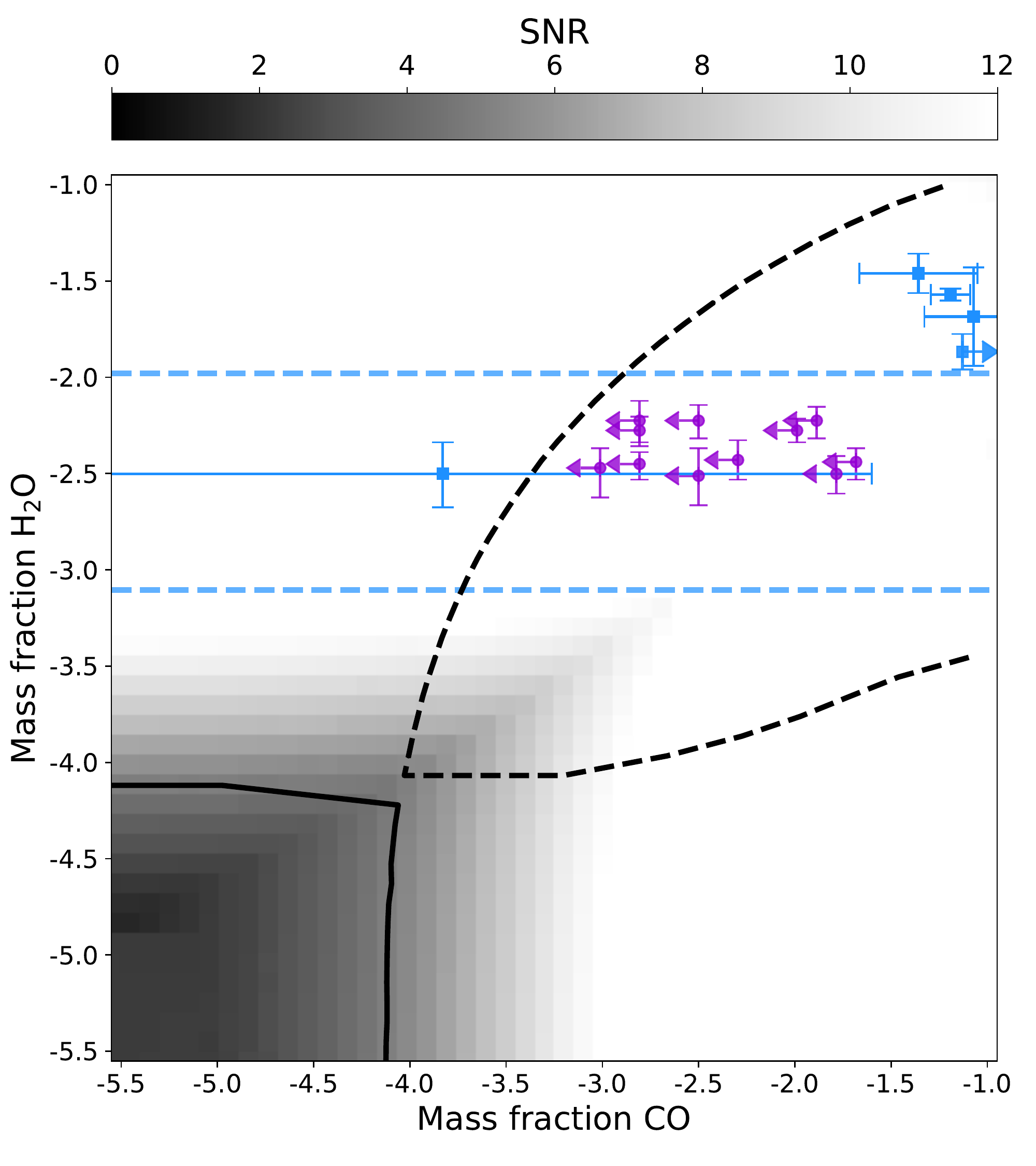}
    \caption{Comparison of the detectability of molecules in the SINFONI data with measured abundances from substellar objects. Blue squares represent directly imaged companions HR8799\,bce, $\beta$\,Pic\,b and 51\,Eri\,b \citep[][submitted]{Lavie2017, Wang2020_HR8799, Molliere2020, Gravity2020, Whiteford2021} while blue lines show the measured range for H$_2$O in $\kappa$\,And\,b \citep{Todorov2016}. The blue arrow indicates the measured value was higher than the range of abundances shown in the plot. Violet circles represent a sample of T-dwarfs analyzed in \cite{Line2017}. In this case arrows refer to upper limits. The solid black line represents the region where no molecule would have been detected in our SINFONI data because of too low mass fractions. Conversely, the dashed black line encloses the region where both molecules would have been detected.}
    \label{fig:comp_abund}
\end{figure}

\begin{figure*}
    \includegraphics[width=1.\hsize]{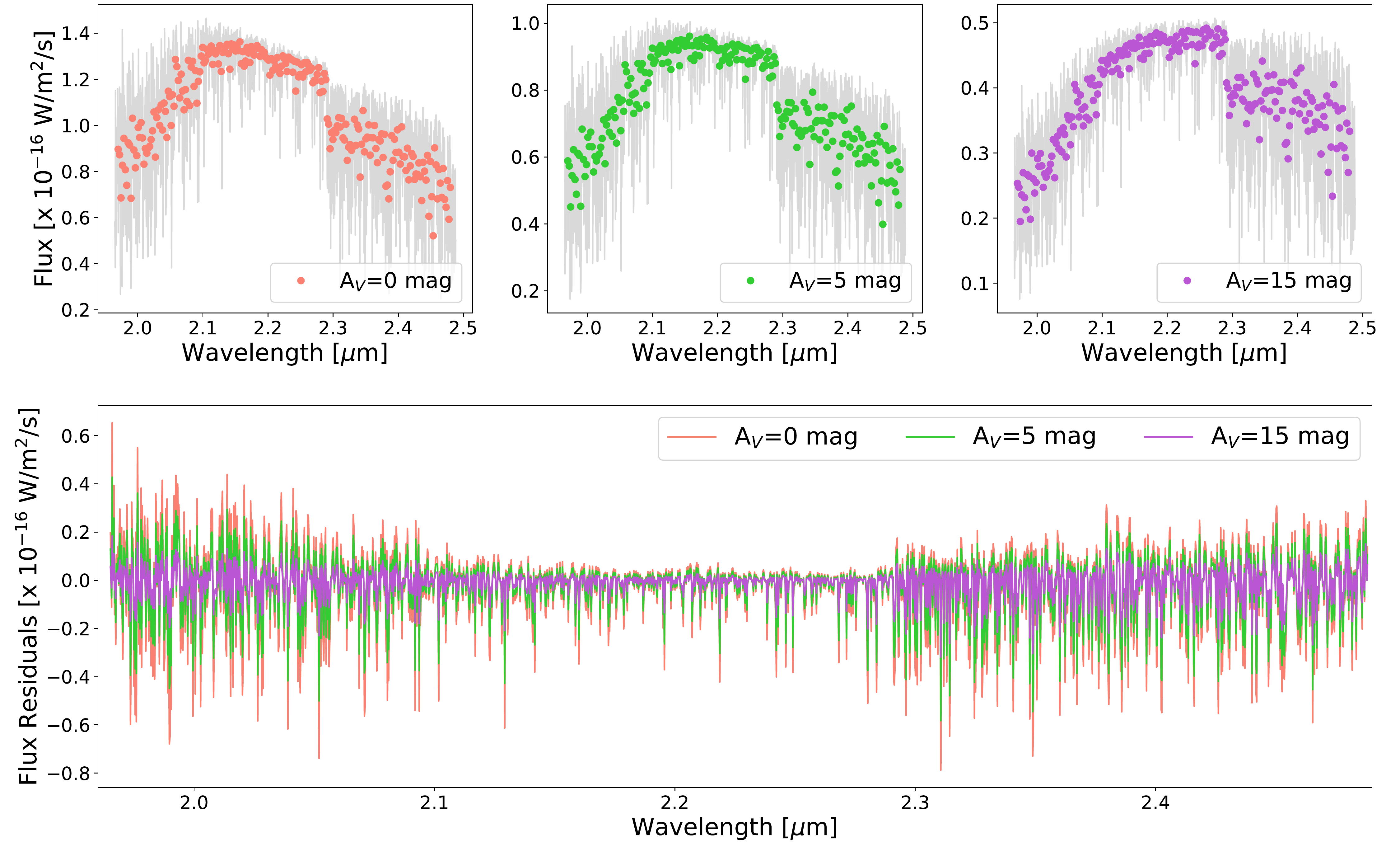}
    \caption{\textit{Top:} Extincted best-fit BT-Settl models resampled at the SINFONI (grey) and at the GRAVITY (colored circles) spectral resolution. Even with high extinction, the broad-band absorption features at the edges of the $K$-band spectrum can be identified. \textit{Bottom:} Spectral features at the SINFONI resolution after the removal of the continuum. Higher extinctions decrease the spectral signatures molecular mapping is looking for, impeding the detection of molecules.}
    \label{fig:SED_vs_MolMap}
\end{figure*}

\begin{figure*}
    %\centering
    \includegraphics[width=1.\hsize]{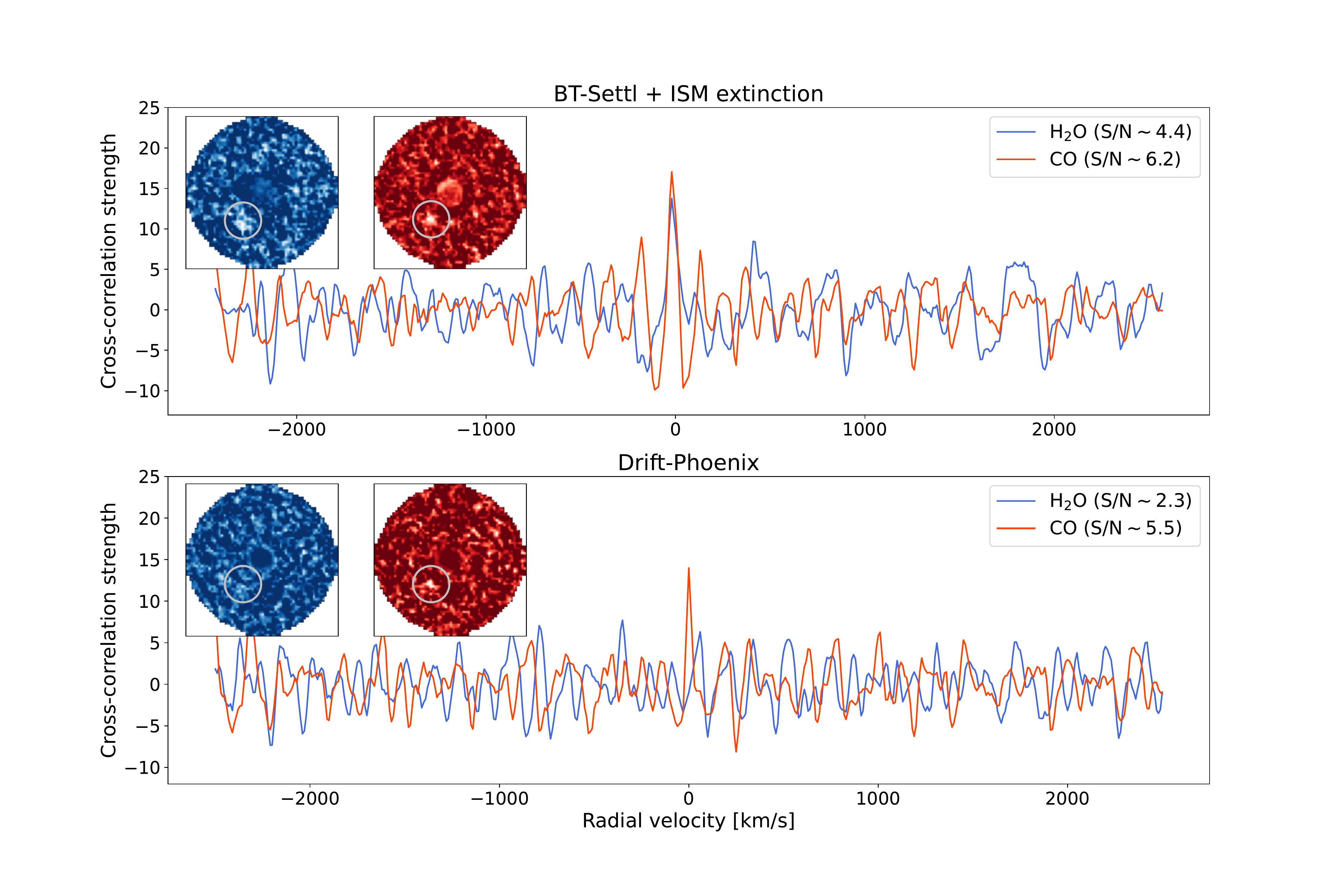}
    \caption{Cross-correlation function for injected planets emitting as the BT-Settl+ISM extinction (top) and DRIFT-PHOENIX (bottom) best fit models proposed in \cite{Wang2021}. CO is detected in both cases, while H$_2$O only in the BT-Settl model. S/N of each detection are reported in the legends. The insets on the top left of each panel show the detection maps for H$_2$O (blue) and CO (red).}
    \label{fig:Wang}
\end{figure*}

\subsection{Chemical comparison with other objects}

We compare PDS70\,b chemical abundance limits with measured values from the literature to put its chemistry in the context of other observed sub-stellar objects. We mainly focus on the chemical abundances, leaving aside the temperature differences between the objects. We injected artificial planets in the data ($\Teff=1200$ K, $\log(g)=3.0$) with different mass fractions of H$_2$O and CO and then tried to detect both individually. The colormap of Fig.~\ref{fig:comp_abund} reports the value of the highest S/N detection among the two molecules. The solid contour line encloses the region where no molecule was detected with S/N>5, while the dashed line highlights the region where both molecules could be identified in the data (i.e., both molecules have S/N>5).

Previous works have exploited the capabilities of atmospheric retrievals \citep{Benneke&Seager2012,Lee2012,Line2012,Line2013,Barstow2013,Barstow2014,Waldmann2015,Waldmann2015b,Heng&Lyons2016} to determine molecular abundances in planet atmospheres, mainly focusing on transiting Hot Jupiters and low-mass field objects. Only a few of the directly imaged planets have been studied with atmospheric retrievals \citep[e.g., HR8799\,bcde and 51\,Eri\,b, ][submitted]{Lee2013, Lavie2017, Wang2020_HR8799,Molliere2020, Whiteford2021} and their abundances are shown with blue squares in Fig~\ref{fig:comp_abund}. 
Furthermore, \cite{Todorov2016} applied atmospheric retrieval on the spectrum of the substellar companion $\kappa$ And b ($\Teff\sim2000$ K), concluding that clouds played a minor role in the atmosphere. They put constraints on the water abundance (shown with a blue dashed line), and a limit on the presence of methane, but were not able to say anything regarding CO. 

\cite{Line2017} applied atmospheric retrieval analysis to 11 late T-dwarfs ($\Teff\sim 600-700$ K) looking for physical/compositional trends. Late T-dwarfs usually show deep spectral features due to H$_2$O absorption which combined with cloud-free atmospheres, allow to probe a wide range of pressures/altitudes. 
They found constraints on H$_2$O, CH$_4$ and NH$_3$, and only upper limits on CO, CO$_2$ and H$_2$S consistent with non-detections (violet circles in Fig.~\ref{fig:comp_abund}). 
Given their water abundances ($\log(\XH2O)\sim-2.6$ to $-2.3$), they show chemical properties different than those from PDS70\,b.

Regarding measured abundances in transiting Hot Jupiters, the large uncertainties \citep[often 2-3 orders of magnitude for H$_2$O and 5-6 for CO, e.g.,][]{Line2014} prevent us from an informed comparison, as the chemical constraints are too loose. For this reason, we do not report them in Fig.~\ref{fig:comp_abund}. Hopefully, future missions like the James Webb Space Telescope \citep[JWST,][]{Gardner2006} and the Atmospheric Remote-sensing Infrared Exoplanet Large-survey \citep[ARIEL,][]{Tinetti2016} will improve measurements of chemical composition of Hot Jupiters and our understanding of their atmospheres.

Assuming we are observing directly the atmosphere of the protoplanet, our analysis suggests that the chemical composition of PDS70\,b is substantially different than those found for other substellar objects whose molecular composition is known, particularly under the assumption of a clear atmosphere.

\subsection[Comparison with previous studies on PDS70 b]{Comparison with previous studies on PDS70\,b}
\label{sec:comparison_previous_studies}
\cite{Wang2021} interpreted the stronger support of the data for atmospheric models than a blackbody spectrum as a proof that we are indeed looking at the atmosphere of the planet. Our VLT/SINFONI data does not support this hypothesis, but we can also not discard it. Dust extinction has a different impact on the detectability of absorption lines than on broad atmospheric features, as shown in Fig.~\ref{fig:SED_vs_MolMap}: with increasing extinction the S/N of absorption lines is expected to get smaller, making a molecular mapping detection more difficult. This is shown in the bottom panel, where for $A_V = 0,5,15$ mag we show the absorption signature of the BT-Settl model reported in \cite{Wang2021} as the best-fit model for PDS70\,b ($\Teff=1392$~K, $\RP=1.96\RJ$, $\log(g)=3.83$). Conversely, dust extinction may also modify the spectral slope at wavelengths where broad features are present, but the triangular $K$-band shape associated with H$_2$O bands remains visible even with $A_V = 15$~mag. The effect can be seen in the three top panels of Fig.~\ref{fig:SED_vs_MolMap}, which display in grey the extincted BT-Settl model at the SINFONI resolution, while colored circles show the spectrum down-sampled to the resolution of VLTI/GRAVITY. It is therefore plausible that dust extinction affects disproportionately line absorption features and broadband features, making molecular mapping in our SINFONI data unsuccessful in detecting molecules in the planet atmosphere. Hence, to better constrain the properties of the dust enshrouding PDS70\,b, higher spectral resolution data able to either detect absorption lines or put tighter constraints on the amount of dust/extinction are required.

In addition to detecting molecular absorption at the blue and red end of the $K$-band, \cite{Wang2021} tested several atmospheric models, augmenting them with dust extinction and possible circumplanetary disk contribution. Here we considered the two models that provided the strongest Bayesian evidence \citep[BT-Settl and DRIFT-PHOENIX;][]{Allard2012, Helling2008}, and for those consider the model augmentation following the same criteria (ISM extinction and no augmentation, respectively). In a first phase, we used those models (without ISM extinction for BT-Settl) to search for cross-correlation signals. Indeed, combining the signature of different molecules present in self-consistent models could boost the S/N and reveal the detection of the atmosphere of the planet. This approach provided successful results in previous works \citep{Hoeijmakers2018,Petrus2020}. Nonetheless, using the pre-computed model grids indicated from the best-fit of the SED did not allow the detection of molecular signals via cross-correlation.
In a second step, we inserted the models in the data as explained in Sect.~\ref{sec:mol_map_det_lim}, and then tried to detect H$_2$O and CO. In both cases, CO is expected to be detected, while H$_2$O only in the BT-Settl + ISM extinction model, even though with a low significance (S/N~$\sim4.5$). The CCF are presented in Fig.~\ref{fig:Wang}, while the insets show molecular maps. Given the confidence of the expected detection (S/N reported in the legend of Fig.~\ref{fig:Wang}) which is inconsistent with the non-detections reported in Sect.~\ref{sec:analysis}, we can exclude that either of the models is able to properly describe the emission coming from PDS70\,b. 

However, even though the pure DRIFT-PHOENIX model maximizes the likelihood, provides a high Bayesian evidence and delivers a good fit to the spectral and photometric data points, the combination of $\log(g)$ and $R_p$ reported in Tab.~4 of \cite{Wang2021} seems to indicate that PDS70\,b has a mass of  $\sim230\MJ$ ($\approx10$--$500~\MJ$ considering the 95\% interval). As pointed out in the same work, such a large mass is in clear contrast with the $10\,\MJ$ upper limit on the planet mass based on orbital fitting enforcing stability \citep{Wang2021}. Hence, although broad H$_2$O features were detected in the VLTI/GRAVITY spectrum and overall dusty models seem to better explain the SED datapoints, none of the so far presented atmospheric models is able to satisfy all the constraints arising from different analysis approaches (orbital fitting, molecular mapping, SED modelling). These self-consistent model grids have proven to be effective in describing more evolved substellar companion atmospheres and are widely used by the community. Nevertheless, we argue that their efficacy in providing a full description of companion emission spectra does not apply to young protoplanets that are still forming like PDS70\,b.

%-----------------------------------------------------------------------

\section{Summary and conclusions}
\label{sec:conclusions}
Medium-resolution (R$\approx$5075) data from PDS70\,b taken with VLT/SINFONI confirmed the complexity of constraining the chemical composition of forming giant planets. We applied molecular mapping to detect absorption features from the planet atmosphere in the $K$-band , without finding any molecular signal.
The non-detection of molecules in the protoplanet atmosphere---despite very deep data in terms of time on target---confirms that a dusty environment likely enshrouds the forming planet. The constraints are not tight enough to yield information about the dust's physical properties. We showed that every self-consistent atmospheric model used in previous works to describe the SED contains spectral features that should have been detected in the SINFONI data. This inconsistency underlines how molecular mapping can provide crucial information to SED fitting with atmospheric models when studying young forming protoplanets, and both techniques need to be explored in more detail to understand the contradicting results.
An approach involving multiple techniques, including molecular mapping, is probably required to combine the different pieces composing planet formation and further constrain the physics and chemistry governing this process. 

Higher resolution data with VLT/ERIS \citep{Davies2018}, VLT/CRIRES+ \citep{Follert2014}, or even with the proposed HiRISE \citep{Vigan2018} might help in shedding light on the presence of molecules and spectral features of the object in the near-infrared, while observations at longer wavelengths, possibly with JWST/MIRI could help in better constraining the emission from the circumplanetary environment. 

%-----------------------------------------------------------------------

\begin{acknowledgements}
We would like to thank the anonymous referee, whose careful and constructive comments improved the quality of this manuscript. GC and SPQ thank the Swiss National Science Foundation for financial support under grant number 200021\_169131. Part of this work has been carried out within the framework of the National Centre of Competence in Research PlanetS supported by the Swiss National Science Foundation. AB and SPQ acknowledge the financial support of the SNSF. PM acknowledges support from the European Research Council under the European Union's Horizon 2020 research and innovation program under grant agreement No. 832428. I.S. acknowledges funding from the European Research Council (ERC) under the European Union's Horizon 2020 research and innovation program under grant agreement No 694513. G-DM acknowledges the support of the DFG priority program SPP 1992 ``Exploring the Diversity of Extrasolar Planets'' (KU~2849/7-1 and MA~9185/1-1) and from the Swiss National Science Foundation under grant BSSGI0$\_$155816 ``PlanetsInTime''. TS acknowledges the support from the ETH Zurich Postdoctoral Fellowship Program and the Netherlands Organisation for Scientific Research (NWO) through grant VI.Veni.202.230. This work has made use of data from the European Space Agency (ESA) mission {\it Gaia} (\url{https://www.cosmos.esa.int/gaia}), processed by the {\it Gaia} Data Processing and Analysis Consortium (DPAC, \url{https://www.cosmos.esa.int/web/gaia/dpac/consortium}). Funding for the DPAC has been provided by national institutions, in particular the institutions participating in the {\it Gaia} Multilateral Agreement. 

\end{acknowledgements}

\bibliographystyle{aa}
\bibliography{PDS70_SINFONI.bib}

\begin{appendix} 
\section{Data selection}
\label{App:Data_selection}
\begin{figure}
    %\centering
    \includegraphics[width=\hsize]{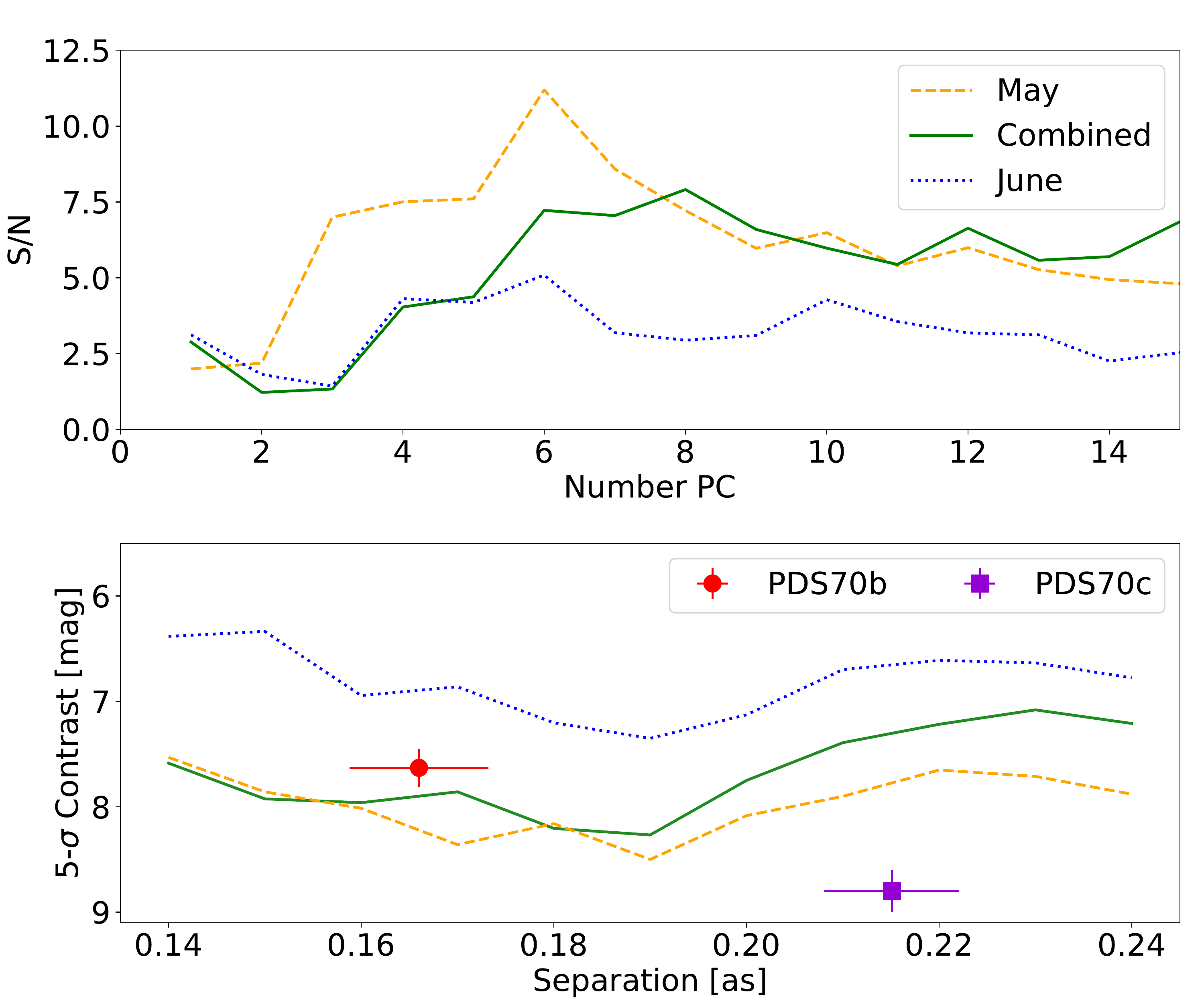}
    \caption{\textit{Top:} S/N of PDS70\,b for the first dataset (orange dashed line), the second one (blue dotted line) and the combination of the two (green solid line) as a function of the number of removed principal components. \textit{Bottom:} Contrast curves obtained for the three considered datasets. Colors and linestyles are the same as in the top panel. PDS70\,b separation and contrast estimates (see App.~\ref{sec:broad_band}) is marked as a red square. PDS70\,c position and contrast are taken from \cite{Haffert2019}. }
    \label{fig:data_selection}
\end{figure}

To verify the data quality, we calculated signal to noise ratios (S/N) and detection limits for the individual observations, as well as for the combination of the two datasets. The proceeding to calculate the S/N follows the prescriptions from \cite{mawet2014} and the detection limits were estimated using PynPoint as explained in \cite{Stolker2020} after the planet was removed from the data (see later). In short, after measuring at a given separation the azimuthal noise and self-subtraction effects on a high S/N injected planet, we estimated the contrast for a false positive fraction of $2.9\times10^{-7}$, corresponding to 5$\sigma$ confidence in the gaussian limit, and including the penalty factor due to small number statistics \citep{mawet2014}.
In the top panel of Fig.~\ref{fig:data_selection} we show the S/N for the first dataset, for the second dataset and for the combination of the two as a function of the removed PCs. The highest S/N is always obtained when looking at the first dataset only (S/N$\sim7.5-12$), and the S/N estimated for the second dataset never reaches values >6. This dataset actually degrades the overall quality when combining with the one from May (green curve). As next step, we generated contrast curves to estimate detection limits for planets detectable with Confidence Level\,>\,5\,$\sigma$. The resulting detection limits, shown in the bottom panel of Fig.~\ref{fig:data_selection}, indicate that with the second dataset it is impossible to detect PDS70\,b (shown as a red circle, see App.~\ref{sec:broad_band} for contrast estimation). When combined with the first dataset, the achieved detection limits are lower than when the first dataset alone is considered. Hence, the first of the two datasets was taken under generally better conditions, leading to a more robust detection of PDS70\,b. 

\begin{figure*}[t]
    \includegraphics[width=1.\hsize]{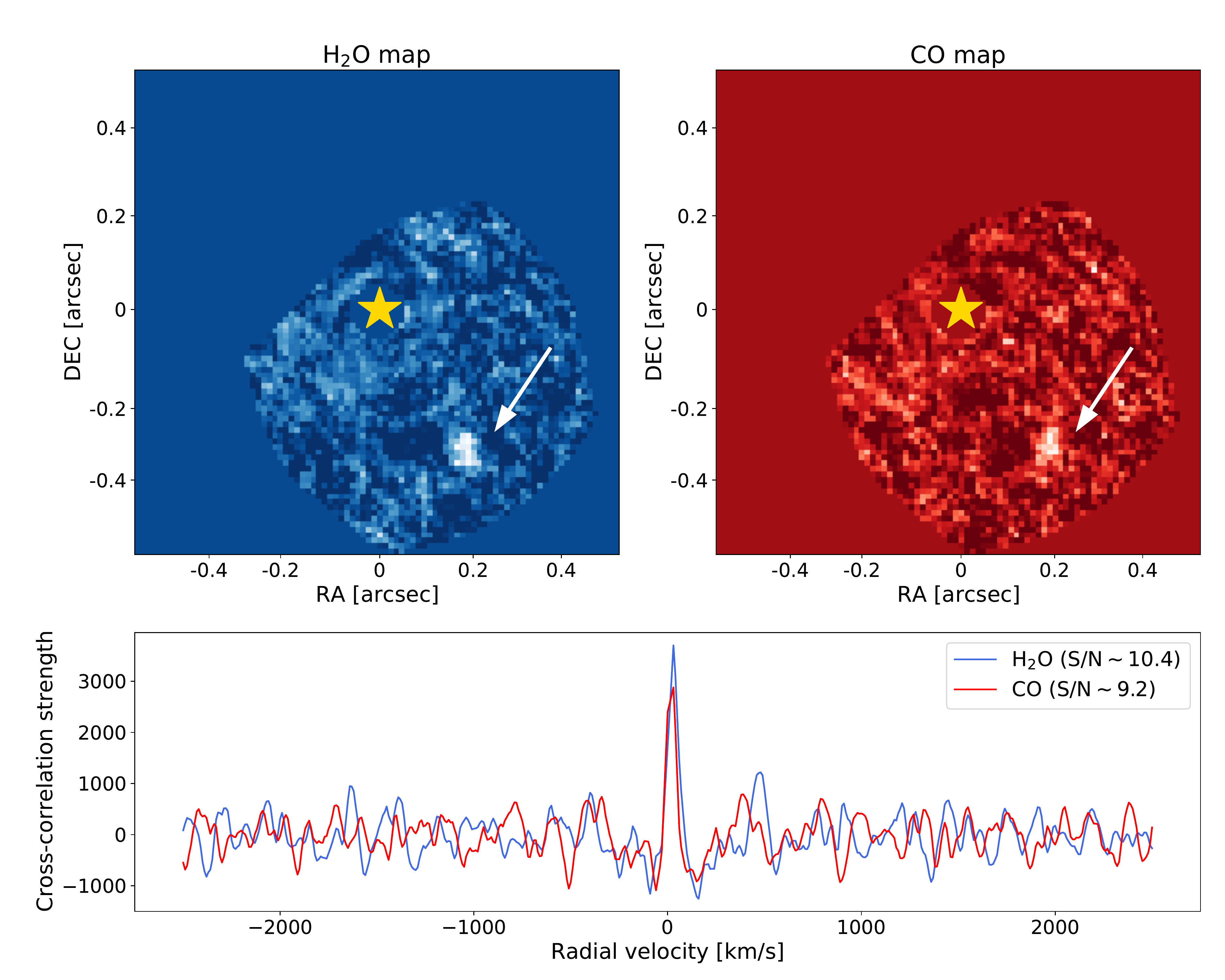}
    \caption{\textit{Top:} Molecular maps for H$_2$O and CO obtained from a dataset of $\beta$\,Pic published in \cite{Hoeijmakers2018}. North is pointing to the top, East to the left. \textit{Bottom:} Cross-correlation function extracted at the planet position. The peaks representing the detection of water and carbon monoxide are clearly visible. }
    \label{fig:Mol_map_BPic}
\end{figure*}

\section{Wavelength-collapsed cubes}
\label{sec:broad_band}
\begin{table}
\caption{Astrometric and photometric values for PDS70b}
\begin{tabular}{l|lll}
\hline\hline
\noalign{\smallskip}
Parameter                     & MCMC chain            & Bias                  & Final value            \\
\noalign{\smallskip}\hline
\noalign{\smallskip}
Separation {[}mas{]}          & $167.4\pm3.3$ & $0.4\pm6.7$   & $167.0\pm7.4$ \\
Position angle {[}$^\circ${]} & $144.7\pm0.6$ & $0.1\pm1.0$ & $144.5\pm1.3$  \\
$\Delta$mag {[}mag{]}         & $7.6\pm0.1$   & $0.0\pm0.2$  & $7.6\pm0.2$\\  \noalign{\smallskip} 
\hline
\end{tabular}
\tablefoot{Final values include instrumental uncertainties (see text).}
\label{tab:astro-photo_PDS70b_wl}
\end{table}

A commonly used method to characterize astrometry and photometry of companions in direct imaging is the injection of negative artificial planets \citep[e.g.][]{Stolker2019, Cugno2019_RCrA}.  This forward modeling technique allows to measure the relative brightness and position of a companion despite self-subtraction, which can undermine direct measurements from the final PSF-subtracted images. Our approach was to iteratively minimize the residuals at the position of the companion after inserting a negative, rescaled copy of the PSF at its location and performing PSF-subtraction with PCA. In our case the PSF template was taken directly from each image, since the images were never saturated.

In order to sample posterior distribution for separation, position angle (PA) and contrast, we used the MCMC algorithm from {\tt PynPoint} presented in \cite{Stolker2019}. We let 200 walkers sample the parameter space with chains of 500 steps. For each walker, the first 100 steps were not considered (burn-in phase). The residuals were minimized in a circular aperture of diameter 2.0 FWHM ($0\farcs11$) centred on the companion position removing 6 PCs at each step, as this is the setting that delivered the highest S/N (see top panel of Fig.~\ref{fig:data_selection}). Since uncertainties delivered by the MCMC algorithm were slightly asymmetric, we conservatively considered the largest of the 2 values. The results of the chain were examined, meaning that a negative signal with the retrieved contrast was injected in the images and after PSF subtraction the images were visually inspected, to verify that no residuals from the companion are visible. The residuals minimization has been effective for all considered cases.

In order to quantify any potential bias in the measurements due to speckles and stellar PSF residuals, we injected artificial point sources at the same separation and with the same contrast as PDS70\,b, at 360 equi-distant PAs \citep{Stolker2020}. The same minimization algorithm was used to retrieve the parameter values and determine possible biases. The resulting offset between the injected and the retrieved values was calculated as the median of the distribution for each parameter. The uncertainties are conservatively estimated to be the maximum between the 16$^{\rm th}$ and 84$^{\rm th}$ percentile of the offset distribution. 

The final values for each parameter was obtained by adding the offset to the value estimated using the MCMC chain. Final uncertainties result from the sum in quadrature of the uncertainties of the MCMC chain and of the bias estimation. Also, uncertainties on the plate scale and position angle obtained by \cite{Meshkat2015} were added in quadrature, although they were estimated for a companion at larger separation \citep{Christiaens2019_ASDI}. These are 0.4 mas in the separation and 0.5$^\circ$ in the PA. All values (and uncertainties) in terms of position and contrast for PDS70\,b are reported in Tab.~\ref{tab:astro-photo_PDS70b_wl}. 

The calculated contrast is in strong agreement with the contrast calculated from IRDIS K1+K2 2016 data from \cite{Keppler2018}, but our value is smaller (meaning the planet is measured to be brighter) than what was estimated by \cite{Muller2018} and \cite{Stolker2020b} for a similar dataset taken in 2018. Finally, our contrast matches the one estimated by \cite{Christiaens2019_ASDI} from SINFONI $H+K$ data. The position that we found is consistent with the past detections of PDS70\,b \citep{Keppler2018, Muller2018, Haffert2019, Wang2020, Stolker2020b}

\section[Molecular mapping on beta Pictoris b]{Molecular mapping on $\beta\,$Pictoris b}
\label{sec:betaPicb}
In this section, we aim at proving that our data reduction and analysis is successfull when molecules in the planet photosphere are abundant enough and their features are not damped by dust. We do this by analyzing a VLT/SINFONI dataset of $\beta$\,Pic already presented in \cite{Hoeijmakers2018}. Compared to their work, we only used data from one of the two nights (09/11/2014, 50 min of time on target) when the star was in the field of view. Indeed, the observation taken on 09/10/2014 (96 min) was carried out with $\beta$~Pic~A outside of the $0\farcs8\times0\farcs8$ field of view and was therefore systematically different than our PDS70 data. We applied the same pipeline used on the PDS70 data (Sect.~\ref{sec:data_reduction}) and cross-correlated the residuals with templates containing H$_2$O and CO (Sect.~\ref{sec:spectral_templates}). The resulting maps are shown in the top panels of Fig.~\ref{fig:Mol_map_BPic}, with the molecular signature from the planet atmosphere correctly identified south-west from the star. The corresponding cross-correlation function is shown in the bottom panel of Fig.~\ref{fig:Mol_map_BPic} for both molecules. We obtained S/N$_{\text{H}_2\text{O}}$ of $\sim10.4$, and S/N$_{\text{CO}}\sim 9.2$, lower than the values reported in \cite{Hoeijmakers2018}, as we only consider one observing night. The analysis on the $\beta$ Pic dataset shows that our pipeline is effective in identifying signals from molecular absorption in planetary atmospheres.

\section{Flux calibration}
\label{sec:flux_calibration}
To estimate the PDS70\,A spectrum we used the observation of the standard star HIP81214 that was executed as part of the calibration. To obtain the spectrum of HIP\,81214 we used the same pipeline applied to the PDS70 data, obtaining a cube covering the spectral range between 2.088\;$\mu$m and 2.44\;$\mu$m. The photometry in each channel was measured in an aperture of radius 2$\lambda$/D for both the HIP\,81214 cube and the median-combination of the centred PDS70 cubes, and we calculated the ratio between these spectra in each channel.
We used BT-NextGen models \citep{Allard1997} to fit HIP\,81214 photometric points from GAIA, Hipparcos and 2MASS \citep{Gaia2018, Hipparcos, Cutri2003} with {\tt species} \citep{Stolker2020}. Temperature and radius obtained from our fit agreed with the values from the GAIA catalogue\footnote{\url{https://gea.esac.esa.int/archive/}}, while from photometric datapoints it was not possible to constrain metallicity and surface gravity. We assumed a constant 100\% transmission, as the instrumental response is taken into account in the calculation of the spectra from the SINFONI data and any effect cancels out when calculating the ratio with the stellar spectra. 
This synthetic spectrum is multiplied by the flux ratio between the HIP\,81214 and the PDS70\,A spectra, in order to obtain PDS70A flux density. We note that this does not correspond to the pure stellar spectrum, as substantial contribution from an inner disk are to be expected \citep{Wang2021}.
Since the night was not photometric and aperture photometry on PDS70\,A shows flux variations of up to 50\%, we scaled the total spectrum of PDS70\,A to the one calculated in \cite{Stolker2020b} in order to have continua that match to each other. Their spectrum was obtained fitting a power-law function to the 2MASS JHK and WISE W1 and W2 magnitudes of the system. This step makes the stellar spectrum consistent with 2MASS $K$-band photometry.

\end{appendix}

\end{document}